\documentclass[reprint,superscriptaddress,footinbib,aps,prd,10pt]{revtex4-1}
\usepackage{graphicx}% Include figure files
\usepackage{dcolumn}% Align table columns on decimal point
\usepackage{bm}% bold math
\usepackage{amssymb,pslatex}
\usepackage{amsmath,marvosym}
\usepackage{braket}
\renewcommand{\d}{\ensuremath{\mathrm{d}}}
\newcommand{\p}{\partial}
\newcommand{\GZ}{\ensuremath{\mathrm{GZ}}}
\newcommand{\gf}{\ensuremath{\mathrm{gf}}}
\newcommand{\YM}{\ensuremath{\mathrm{YM}}}

\begin{document}

\title{{\Large {\bf BRST-Symmetry Breaking and Bose-Ghost Propagator \\[2mm]
                    in Lattice Minimal Landau Gauge}}}

\author{Attilio~Cucchieri}
\email{attilio@ifsc.usp.br}
\affiliation{Instituto de F\'\i sica de S\~ao Carlos, Universidade de S\~ao Paulo, Caixa Postal 369, 13560-970 S\~ao Carlos, SP, Brazil}
\author{David~Dudal}
\email{david.dudal@ugent.be}
\affiliation{Ghent University, Department of Physics and Astronomy, Krijgslaan 281-S9, 9000 Gent, Belgium}
\author{Tereza~Mendes}
\email{mendes@ifsc.usp.br}
\affiliation{Instituto de F\'\i sica de S\~ao Carlos, Universidade de S\~ao Paulo, Caixa Postal 369, 13560-970 S\~ao Carlos, SP, Brazil}
\author{Nele~Vandersickel$\,$}
\email{nele.vandersickel@ugent.be}
\affiliation{Ghent University, Department of Physics and Astronomy, Krijgslaan 281-S9, 9000 Gent, Belgium}

\begin{abstract}
The Bose-ghost propagator has been proposed as a carrier
of the confining force in Yang-Mills theories in minimal Landau gauge.
We present the first numerical evaluation of this propagator, using
lattice simulations for the SU(2) gauge group in the scaling region.
Our data are well described by a simple fitting function, which is compatible 
with an infrared-enhanced Bose-ghost propagator. This function can also 
be related to a massive gluon propagator in combination with an infrared-free
(Faddeev-Popov) ghost propagator. Since the Bose-ghost propagator can
be written as the vacuum expectation value of a BRST-exact
quantity and should therefore vanish in a BRST-invariant theory,
our results provide the first numerical manifestation
of BRST-symmetry breaking due to restriction of
gauge-configuration space to the Gribov region.
\end{abstract}

\maketitle

%%%%%%%%%%%%%%%%%%%%%%%%%%%%%%%%%%%%%%%%%%%%%%%%%%%%%%%%%%%%%%%%%%%%%%%%%%%%%%%

The study of color confinement in Yang-Mills theories in minimal Landau
gauge is an active area of research \cite{Greensite:2011zz}. Let us
recall that, in this case, the gauge condition is implemented (see for
example \cite{Vandersickel:2012tz} and references therein) by restricting
the functional integral over gauge-field configurations to the so-called
Gribov region $\Omega$. This restriction can be achieved by adding a
nonlocal term $S_{\mathrm{h}}$, the horizon function, to the
usual Landau gauge-fixed Yang-Mills action $S_{\YM} + S_{\gf}$. One thus
obtains the Gribov-Zwanziger (GZ) action $S_{\GZ} = S_{\YM} + S_{\gf}
+ \gamma^4 S_{\mathrm{h}}$. The massive parameter $\gamma$, known as the
Gribov parameter, is dynamically determined (in a self-consistent way)
through the so-called horizon condition.

%%%%%%%%%%%%%%%%%%%%%%%%%%%%%%%%%%%%%%%%%%%%%%%%%%%%%%%%%%%%%%%%%%%%%%%%%%%%%%%

\vskip 3mm

In order to localize the GZ action \cite{Vandersickel:2012tz}
one introduces a pair of
complex-conjugate bosonic fields $(\overline{\phi}^{ac}_{\mu},
\phi^{ac}_{\mu} )$ and a pair of Grassmann complex-conjugate fields
$(\overline{\omega}^{ac}_{\mu}, \omega^{ac}_{\mu} )$. Then, the GZ action
can be written as $ S_{\GZ} = S_{\YM} + S_{\gf} + S_{\mathrm{aux}}
+ S_{\gamma}$, where
\begin{eqnarray}
\!\!\!\!\!\!\!\!\!\! S_{\mathrm{aux}} & = & \int \d^{\rm 4} x \,
\Bigl[ \overline{\phi}_{\mu}^{ac} \, \p_{\nu} \left( D_{\nu}^{ab}
                     \phi^{bc}_{\mu} \right) -
     \overline{\omega}_{\mu}^{ac} \, \p_{\nu} \left( D_{\nu}^{ab}
                     \omega^{bc}_{\mu} \right) \nonumber \\[0mm]
            &   & \;\;\; \qquad \qquad - g_0 \left( \p_{\nu}
         \overline{\omega}_{\mu}^{ac} \right) f^{abd} \,
                D_\nu^{be} \eta^e \phi_{\mu}^{dc} \Bigr] \\[1mm]
\!\!\!\!\!\!\!\!\!\! S_{\gamma} & = & \int \d^{\rm 4}x \,
           \Bigl[ \gamma^{2}
                 D^{ba}_{\nu} \Bigl( \phi_{\nu}^{ab}
           + \overline{\phi}_{\nu}^{ab} \Bigr)
        - 4 \left( N_c^{2} - 1 \right) \gamma^{4} \Bigr] \, .
\label{eq:Sofgamma}
\end{eqnarray}
Here, $a,\,b,\,c,\,d$ and $e$ are color indices in the adjoint
representation of the SU($N_c$) gauge group, $\mu$ and $\nu$ are
Lorentz indices. Also, $g_0$ is the bare coupling constant,
(${\overline \eta}^b$, $\eta^b$) are the Faddeev-Popov (FP) ghost
fields, $D^{ab}_{\nu} = \delta^{ab} \partial_{\nu} + 
g_0 f^{acb} A_{\nu}^{c}$ is the covariant derivative, $f^{abd}$
are the structure constants of the gauge group and
repeated indices are always implicitly summed over.
At the classical level, the total derivatives
$\p_{\nu} ( \phi_{\nu}^{aa} + \overline{ \phi}_{\nu}^{aa} )$ in
the definition of $S_{\gamma}$ can be neglected \cite{Vandersickel:2012tz,
Zwanziger:2009je}.

Under the nilpotent BRST variation $s$ \cite{Baulieu:1983tg}, the four
auxiliary fields form two BRST doublets, i.e.\ $s \, \phi^{ac}_{\mu} =
\omega^{ac}_{\mu}$, $s \, \omega^{ac}_{\mu} = 0$, $s \,
\overline{\omega}^{ac}_{\mu} = \overline{\phi}^{ac}_{\mu}$
and $s \, \overline{\phi}^{ac}_{\mu}= 0$, giving rise to a
BRST quartet. At the same time, one can check that the
localized GZ theory is not BRST-invariant. Indeed, while
$s \, (S_{\YM} + S_{\gf} + S_{\mathrm{aux}}) = 0$, one finds that
$s \, S_{\gamma} \propto \gamma^2 \neq 0$. Since
a nonzero value for the Gribov parameter $\gamma$ is related to the
restriction of the functional integration to the Gribov region
$\Omega$, it is clear that BRST-symmetry breaking is a direct
consequence of this restriction, as investigated in several works
(see e.g.\ \cite{Zwanziger:2009je,
brst,Sorella:2009vt,Zwanziger:2010iz} and references therein).

%%%%%%%%%%%%%%%%%%%%%%%%%%%%%%%%%%%%%%%%%%%%%%%%%%%%%%%%%%%%%%%%%%%%%%%%%%%%%%%

\vskip 3mm

In order to study numerically the effect of the BRST-breaking
term $S_{\gamma}$, one can consider the expectation value of a
BRST-exact quantity. One such possibility is the correlation
function
\begin{equation}
Q^{abcd}_{\mu \nu}(x,y) \, = \,
\braket{ \,
 \omega^{ab}_{\mu}(x) \, \overline{\omega}^{cd}_{\nu}(y) \, + \,
 \phi^{ab}_{\mu}(x) \, \overline{\phi}^{cd}_{\nu}(y) \, } \, ,
\end{equation}
which can be written as $\braket{ \, s (\, \phi^{ab}_{\mu}(x) \,
\overline{\omega}^{cd}_{\nu}(y)) \, }$. Of course, while the above 
expectation value should be zero for a BRST-invariant theory, it does
not necessarily vanish if BRST symmetry is broken
(see, for example, the discussion in Ref.\ \cite{Sorella:2009vt}). 
Indeed, at tree level (and in momentum
space) one finds \cite{Vandersickel:2012tz,Gracey:2010df}
\begin{equation}
\label{eq:Qabcd}
Q^{abcd}_{\mu \nu}(p,p') \, = \,
 \frac{ \left(2 \pi\right)^4 \delta^{(4)}\left(p + p'\right)
        \, g_0^2 \, \gamma^4 f^{abe} f^{cde} P_{\mu \nu}(p)}{
          p^2 \, \left(p^4 + 2 g_0^2 N_c \gamma^4 \right)} \, ,
\end{equation}
where $P_{\mu \nu}(p)$ is the usual transverse projector. Thus,
this propagator is proportional to the Gribov parameter $\gamma$,
i.e.\ its nonzero value is clearly related to the breaking of the
BRST symmetry in the GZ theory.

%%%%%%%%%%%%%%%%%%%%%%%%%%%%%%%%%%%%%%%%%%%%%%%%%%%%%%%%%%%%%%%%%%%%%%%%%%%%%%%

\vskip 3mm

On the lattice one does not have direct access to the auxiliary
fields $(\overline{\phi}^{ac}_{\mu}, \phi^{ac}_{\mu} )$ and
$(\overline{\omega}^{ac}_{\mu}, \omega^{ac}_{\mu} )$. 
On the other hand, by 1) adding suitable sources to the GZ action,
2) explicitly integrating over the four auxiliary fields ---which
enter the action at most quadratically--- and 3) taking the usual
functional derivatives with respect to the sources, one can verify
that \cite{Zwanziger:2009je}
\begin{equation}
\!\!\!\!\!\!\!\!
Q^{abcd}_{\mu \nu}(x-y) \, = \, \gamma^4 \, \left\langle \,
       R^{a b}_{\mu}(x) \, R^{c d}_{\nu}(y) \, \right\rangle \, ,
\label{eq:Qprop}
\end{equation}
where
\begin{equation}
R^{a c}_{\mu}(x) = \int \d^{\rm 4} z \,
         ( {\cal M}^{-1} )^{ae}(x,z) \, B^{ec}_{\mu}(z)
\label{eq:Rfunc}
\end{equation}
and $B^{bc}_{\nu}(x) = g_0 \, f^{b e c} \, A^{e}_{\nu}(x)$.
We remark that the notations
used in Refs.\ \cite{Vandersickel:2012tz} and \cite{Zwanziger:2009je}
are slightly different. Also, our $Q^{abcd}_{\mu \nu}(x,y)$ propagator
corresponds to the $F$-term of the $V$-propagator in equations
(72) and (75) of Ref.\ \cite{Zwanziger:2009je}.
Thus, the behavior of this propagator depends only on the bosonic fields
$(\overline{ \phi}^{ac}_{\mu}, \phi^{ac}_{\mu} )$. Finally, one should
recall that this Bose-ghost propagator has been proposed as a carrier
of long-range confining force in minimal Landau gauge
\cite{Zwanziger:2009je,Furui:2009nj,Zwanziger:2010iz}.

%%%%%%%%%%%%%%%%%%%%%%%%%%%%%%%%%%%%%%%%%%%%%%%%%%%%%%%%%%%%%%%%%%%%%%%%%%%%%%%

\vskip 3mm

We evaluate the Bose-ghost propagator as defined in
Eq.\ (\ref{eq:Qprop}) above ---modulo the global factor
$\gamma^4$--- using numerical simulations in the SU(2) case.
In order to check
discretization effects, we considered three different values
of the lattice coupling $\beta$, i.e.\ $\beta = 2.2$, $2.34940204$
and $2.43668228$, respectively corresponding \cite{Bloch:2003sk} to
a lattice spacing $a$ of about $0.210 \, fm$, $0.140\, fm$ and $0.105
\, fm$. For $\beta = 2.2$ and $2.34940204$ we used five different
lattice volumes, i.e.\ $V = 16^4$, $24^4$, $32^4$, $40^4$, $48^4$ in
the former case and $V = 24^4$, $36^4$, $48^4$, $60^4$, $72^4$ in the
latter case. These two sets yield (approximately) the same set of
physical volumes, ranging from about $(3.366 \, fm)^4$ to $(10.097
\, fm)^4$. For $\beta = 2.43668228$ we considered only the lattice
volume $V = 96^4$, which also corresponds to a physical volume of
about $(10.097 \, fm)^4$. Thermalized configurations have been
gauge-fixed to lattice minimal Landau gauge using the
stochastic-overrelaxation algorithm \cite{Cucchieri:1995pn}, with a stopping criterion
$( \partial_{\mu} \vec{A}_{\mu} )^2 \leq 10^{-14}$ (after averaging
over the lattice volume and the three color components).

In order to evaluate the Bose-ghost propagator, we invert the FP
matrix ${\cal M}^{a b}(x,y)$ for the sources $B^{bc}_{\mu}(x)$,
after removing their zero modes. As for the lattice
gauge field $A_{\mu}(x)$, corresponding to $a g_0 A_{\mu}(x)$ in
the continuum, we employ the usual unimproved definition $[U_{\mu}(x)
- U^{\dagger}_{\mu}(x)]/(2 i)$, where $U_{\mu}(x)$ are the lattice
link variables entering the Wilson action. The inversion of the
FP matrix is performed using a conjugate-gradient method, accelerated
by even/odd preconditioning. If we indicate with
$ \widetilde{R}^{a c}_{\mu}(k) = V^{-1/2} \sum_x R^{a c}_{\mu}(x) \,
\exp{(2 \pi i k \cdot x / N)} $ the Fourier transform of the outcome
$R^{a c}_{\mu}(x)$ of the numerical inversion [see Eq.\
(\ref{eq:Rfunc})], then it is clear that we can evaluate the
Bose-ghost propagator [see Eq.\ (\ref{eq:Qprop})] in momentum space by
considering $\, Q^{abcd}_{\mu \nu}(k) \equiv \Re \{ \,
\widetilde{R}^{a b}_{\mu}(k) \widetilde{R}^{c d}_{\nu}(-k) \,
\}$.
In the above equations, $N$ is the lattice side, $k$ is the
wave vector with components $k_{\mu} = 0, 1, \ldots,
N-1$ and $\Re$ indicates the real part of the expression
within brackets. Then, by contracting the $b$ and $d$ indices,
we can write [see Eq.\ (\ref{eq:Qabcd})]
\begin{equation}
Q^{a c}(k) \, \equiv \, Q^{abcb}_{\mu \mu}(k) \, \equiv \,
    \delta^{a c} N_c \, P_{\mu \mu}(k) \, Q(k^2) \, ,
\label{eq:Q}
\end{equation}
due to global color invariance.

\begin{figure}[t]
\centering
\vskip -2mm
\includegraphics[trim=55 0 40 0, clip, scale=1.00]{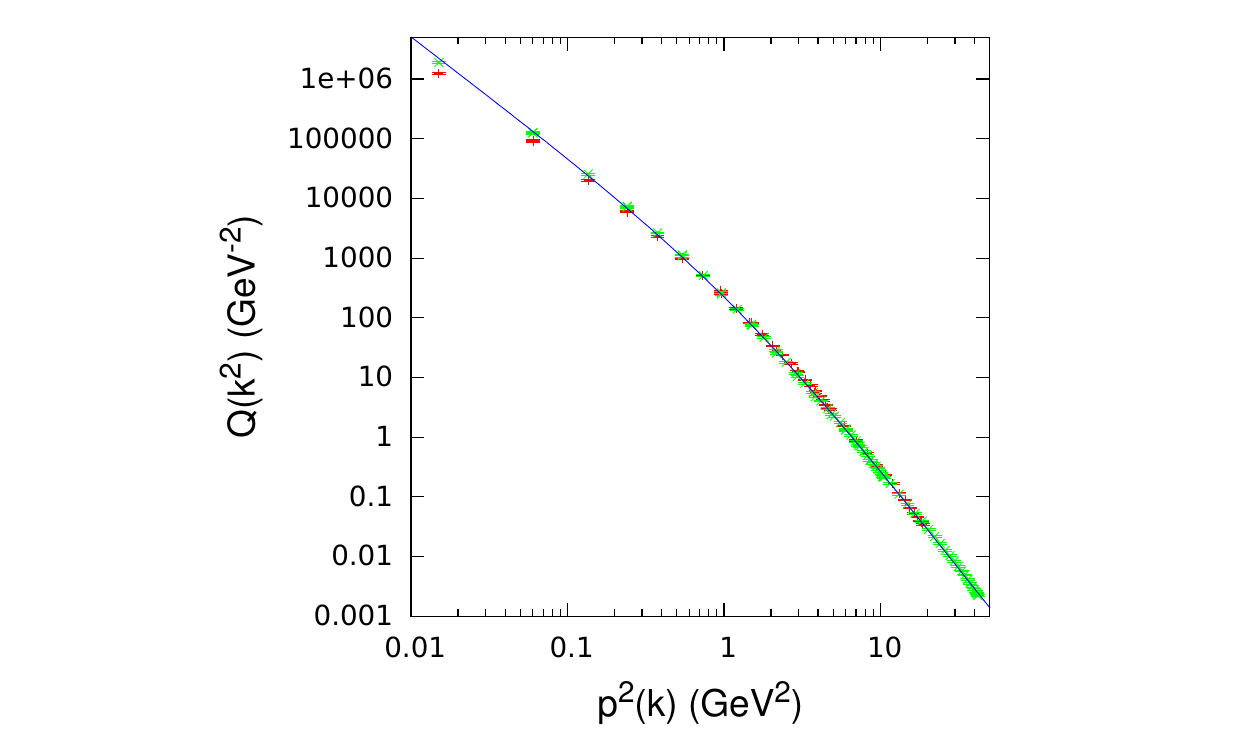}
\caption{\label{fig:fit} The Bose-ghost propagator $Q(k^2)$,
defined in Eq.\ (\ref{eq:Q}), as a function of the improved 
momentum squared $p^2(k)$.
We plot data for $\beta = 2.2$,
$V = 48^4$ (red, $+$, 500 configurations) and $\beta =
2.34940204$, $V = 72^4$ (green, $\times$, 250 configurations),
after applying a matching procedure \cite{matching}. 
We also plot, for $V = 72^4$, 
a fit using Eq.\ (\ref{eq:fit}) and the parameters in Table \ref{tab:fits}, 
with $c = 114(13)$.  Note the logarithmic scale on both axes.
}
\end{figure}

%%%%%%%%%%%%%%%%%%%%%%%%%%%%%%%%%%%%%%%%%%%%%%%%%%%%%%%%%%%%%%%%%%%%%%%%%%%%%%%

\vskip 3mm

\begin{figure}[t]
\centering
\vskip -2mm
\includegraphics[trim=55 0 40 0, clip, scale=1.00]{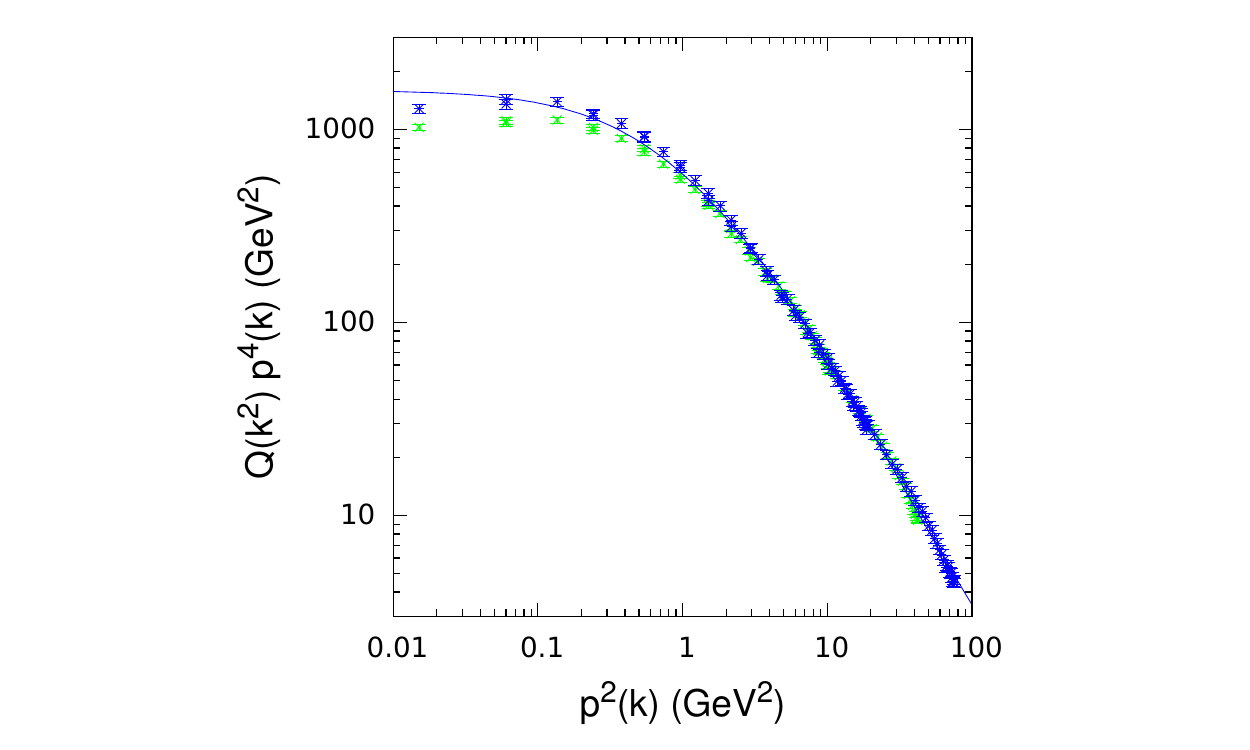}
\caption{\label{fig:fit2} The product $Q(k^2) p^4(k)$, as a function
of the improved momentum squared $p^2(k)$.
We plot data for $\beta = 2.34940204$,
$V = 72^4$ (green, $\times$, 250 configurations) and
$\beta = 2.43668228$, $V = 96^4$ (blue, $*$, 100 configurations),
after applying a matching procedure \cite{matching} to
the former set of data.
We also plot, for $V = 96^4$, 
a fit using Eq.\ (\ref{eq:fit}) and the parameters in Table \ref{tab:fits}, 
with $c = 247(16)$. Note the logarithmic scale on both axes.
}
\end{figure}

Numerical results for the scalar function $Q(k^2)$, defined in
the above equation, are shown in Figs.\ \ref{fig:fit} and
\ref{fig:fit2}. In all cases the data points represent averages
over gauge configurations and error bars correspond to one
standard deviation. (We consider the statistical error only;
the number of configurations ranges from 10000, for $V = 16^4$
at $\beta = 2.2$, to 100, for $V = 96^4$ at $\beta = 2.43668228$.)
In the plots, all quantities are in physical units and we use
the improved definition for the momenta, i.e.\ $p^2(k) = \sum_{\mu}
( p_{\mu}^2 + p_{\mu}^4 / 12 )$ with $p_{\mu} = 2 \sin{(\pi
k_{\mu} / N)}$, which makes the behavior of the propagator smoother,
allowing a better fit to the data. In our simulations we
considered two types of momenta, i.e.\ wave vectors
whose components are $(0,0,0,k)$ and $(k,k,k,k)$, with $k = 1,
2, \ldots, N/2-1$. This gives $N-2$ different values for the
momentum $p$. [Note that the null momentum trivially gives
a zero result for the scalar function $Q(k^2)$, since
$\widetilde{R}^{a c}_{\mu}(0) = 0$.] We also checked that the
extrapolation to the infinite-volume limit is relevant only to
clarify the infrared (IR) behavior of the propagator, i.e.\
finite-size effects ---at a given lattice momentum $p$--- are
essentially negligible. Here, we did not check for possible
Gribov-copy effects.

In Fig.\ \ref{fig:fit} we show the data at $\beta = 2.2$
with $V = 48^4$ and at $\beta = 2.34940204$ with $V = 72^4$,
after rescaling the data at $\beta = 2.2$ using the
matching technique described in Ref.\ \cite{matching}. The
data scale quite well, even though small deviations are observable
in the IR limit (see also Fig.\ \ref{fig:fit2}). We also fit the data
using the fitting function
\begin{equation}
\label{eq:fit}
f(p^2) \, = \, \frac{c}{p^4} \, \frac{p^2 + s}{p^4 \, + \,
                           u^2 p^2 \, + \, t^2 } \, .
\end{equation}
Following the analysis in \cite{Zwanziger:2009je,Zwanziger:2010iz},
i.e.\ by using the relation (obtained using a cluster decomposition)
\begin{equation}
Q(p^2) \, \sim \, g_0^2 \, G^2(p^2) \, D(p^2) \, ,
\label{eq:cluster}
\end{equation}
where $D(p^2)$ is the gluon propagator and $G(p^2)$ is the ghost propagator,
the above fitting function corresponds to considering an infrared-free
ghost propagator $G(p^2)$ and a massive gluon propagator $D(p^2)$
\cite{massive}.
The fit describes the data quite well (see the $\chi^2/\mbox{dof}$
values in Table \ref{tab:fits}). Let us note that the fitting
value for the parameter $c$ is somewhat arbitrary, since
one can always fix a renormalization condition \cite{foot1} 
at a given
scale $p^2 = \mu^2$, which in turn yields a rescaling of the
Bose-ghost propagator by a global factor. 
\setlength{\tabcolsep}{3.5pt}
\begin{table}[t]
\begin{tabular}{| c | c | c | c | c | c |}
\hline
$V = N^4$      & $\beta$    & $t \, (GeV^2)$ & $u \, (GeV)$ &
$s \, (GeV^2)$ & $\chi^2 / \mbox{dof}$ \\
\hline
$\,48^4\,$ & 2.2        & 2.2(0.2) & 1.5(0.2) & 9.9(3.1) &
6.28 \\ \hline
$\,72^4\,$ & 2.34940204 & 3.2(0.3) & 3.6(0.4) & 46(13) &
2.40 \\ \hline
$\,96^4\,$ & 2.43668228 & 3.0(0.2) & 3.9(0.3) & 58.0(9.8) &
1.12 \\ \hline
\end{tabular}
\caption{
Parameters $t$, $u$ and $s$ from a fit of 
$f(x)$ in (\ref{eq:fit}) to the data. 
Errors in parentheses correspond to one standard
deviation. The number of degrees of freedom (dof) is always $N-6$. 
We also show the reduced chi-squared $\chi^2 / \mbox{dof}$. 
Fits have been done using {\tt gnuplot}.
\label{tab:fits}}
\end{table}

On the other hand, the
parameters $t$, $u$ and $s$ can be related to the analytic structure of
the Bose-ghost propagator $Q(p^2)$. For example, as done in Ref.\
\cite{Cucchieri:2011ig} for the gluon propagator, one could try
to rewrite the fitting function in terms of a pair of complex-conjugate
poles. Then, we find that these poles are actually real and given
by $ 7.6(1.2) \pm 7.0(1.3) $, where we used the data reported in
the last line of Table \ref{tab:fits}. (Errors, shown in parentheses,
correspond to one standard deviation and were obtained using
propagation of error.) Thus, this fit supports the so-called massive
solution of the coupled Yang-Mills Dyson-Schwinger equations of gluon
and ghost propagators (see e.g.\ Ref.\ \cite{Aguilar:2008xm})
and the so-called Refined GZ approach \cite{RGZ}. However,
the values for the fitting parameters do not seem to relate in a simple
way to the corresponding values obtained by fitting gluon-propagator
data \cite{Cucchieri:2011ig}.

Even though the simple Ansatz above gives a good description of
the data, deviations can be seen in the IR region for momenta
below about $1 \, GeV$, by plotting the quantity $Q(k^2) \, p^4(k)$
(see Fig.\ \ref{fig:fit2}). We checked that one can slightly
improve our fits, by using more general forms of the propagator.
However, in these cases, most of the fitting parameters are determined
with very large errors, suggesting that such fitting functions have too
many (redundant) parameters.
For this reason we do not show these fits here.
(A more detailed analysis of the lattice data will be
presented elsewhere.)

%%%%%%%%%%%%%%%%%%%%%%%%%%%%%%%%%%%%%%%%%%%%%%%%%%%%%%%%%%%%%%%%%%%%%%%%%%%%%%%

\vskip 3mm

The above results allow a simple qualitative description of
the momentum-space behavior of the Bose-ghost propagator. In
particular, we find that its IR behavior is strongly enhanced,
given by $p^{-4}$. This result is in agreement with the one-loop
analysis carried out in \cite{Gracey:2009mj} but not with the
prediction of Ref.\ \cite{Zwanziger:2009je}, where an IR-enhancement
of $p^{-6}$ was obtained by considering in Eq.\ (\ref{eq:cluster})
an IR-enhanced ghost propagator and an IR-vanishing gluon
propagator (see for example Refs.\ \cite{scaling}).
One should stress that, even though a double-pole singularity is
suggestive of a long-range interaction, the above result does not
imply a linearly-rising potential between quarks. Indeed, when
coupled to quarks via the $A-\phi$ propagator ---which is nonzero
due to the vertex term $\overline{\phi}_{\mu}^{ac} \, g f^{acb}
A_{\nu}^{c} \p_{\nu} \, \phi^{bc}_{\mu}$---, the Bose-ghost
propagator gets a momentum factor at each vertex
\cite{Zwanziger:2009je,Zwanziger:2010iz}, i.e.\ the effective
propagator is given by $p^{-2}$ in the IR limit. This analysis
is confirmed by the explicit evaluation of the static potential
in Ref.\ \cite{Gracey:2009mj}, considering a two-loop topology
with the exchange of a Bose-ghost quantum. These results seem
to suggest that a linearly rising potential cannot be obtained
by a perturbative calculation based on a simple one-particle exchange,
but requires a fully non-perturbative analysis. (For a
lengthier discussion about this issue, the reader can refer, for
example, to the last section in Ref.\ \cite{Gracey:2009mj}.)

%%%%%%%%%%%%%%%%%%%%%%%%%%%%%%%%%%%%%%%%%%%%%%%%%%%%%%%%%%%%%%%%%%%%%%%%%%%%%%%

We conclude by stressing that, even though we did not explicitly
evaluate the Gribov parameter $\gamma$ \cite{foot2},
our results constitute the first numerical manifestation of 
BRST-symmetry breaking due to the restriction of the functional 
integration to the Gribov region $\Omega$ in the GZ approach.
This directly affects continuum functional approaches in
Landau gauge (see for example \cite{bse} and references therein), which
usually employ lattice results in minimal Landau gauge as an input and/or
as a comparison. At the same time, as stressed in Section 8 of the recent
QCD review \cite{Brambilla:2014aaa}, several questions are still open for
a clear understanding, at the nonperturbative level, of the GZ approach.
In particular, one should understand how a physical positive-definite
Hilbert space could be defined in this case.

%%%%%%%%%%%%%%%%%%%%%%%%%%%%%%%%%%%%%%%%%%%%%%%%%%%%%%%%%%%%%%%%%%%%%%%%%%%%%%%

\vskip 3mm

\noindent
{\bf Acknowledgments:}
The authors thank S.P.\ Sorella for useful discussions. A.~Cucchieri and
T.~Mendes acknowledge partial support from FAPESP ({\bf grant \#
2009/50180-0}) and from CNPq. 
D.~Dudal and N.~Vandersickel are supported by the Research
Foundation-Flanders.
We also would like to acknowledge
computing time provided on the Blue Gene/P supercomputer
supported by the Research Computing Support Group (Rice
University) and Laborat\'orio de Computa\c c\~ao Cient\'\i fica
Avan\c cada (Universidade de S\~ao Paulo).

%%%%%%%%%%%%%%%%%%%%%%%%%%%%%%%%%%%%%%%%%%%%%%%%%%%%%%%%%%%%%%%%%%%%%%%%%%%%%%%

%%%%%%%%%%%%%%%%%%%%%%%%%%%%%%%%%%%%%%%%%%%%%%%%%%%%%%%%%%%%%%%%%%%%%%%%%%%%%%%


\begin{thebibliography}{99}
\bibitem{foot2}
One should recall
that the parameter $\gamma$ is not explicitly introduced on the lattice, 
since the restriction of the gauge-configuration space to the
region $\Omega$ is achieved by numerical minimization. Thus, quantities
proportional to $\gamma$, such as the Bose-ghost propagator considered
here or the horizon function (see, e.g., \cite{Cucchieri:1997ns}), are
always evaluated modulo the global $\gamma^4$ factor.
\end{thebibliography}
\end{document}